# Electronic structure by X-ray absorption spectroscopy and observation offield induced unusually slowspin relaxation from magnetic properties in pyrochlore$Eu_{2-x}Fe_xTi_2O_7$

ArkadebPal[1],Surajit Ghosh[1], Shiv Kumar[2], Eike F. Schwier[2], Masahiro Sawada[2], Kenya Shimada[2],Mukul Gupta[4],D. M. Phase[4], A. K. Ghosh[3] and Sandip Chatterjee[1,*]

[1] Department of Physics, Indian Institute of Technology (Banaras Hindu University), Varanasi-221005, India

[2]Hiroshima Synchrotron Radiation Center, Hiroshima University, Kagamiyama 2-313, Higashi-Hiroshima 739-0046, Japan

[3]Department of Physics, Banaras Hindu University, Varanasi-221005, India

[4]UGC-DAE Consortium for Scientific Research, University Campus, KhandwaRoad, Indore-452001, India

*Corresponding author e-mail id: schatterji.app@iitbhu.ac.in

## Abstract

X-ray absorption spectroscopy (XAS) as well as x-ray magnetic circular dichroism (XMCD) and magnetization of hybrid pyrochlore $Eu_{2-x}Fe_xTi_2O_7$ were investigated, where the rare earth Eu (4*f*) was replaced with transition metal Fe (3*d*) to introduce competing 4*f*-3*d* interactions. It is confirmed that the valence states of Eu and Fe ions are formally trivalent while that of Ti ions are tetravalent ($3d^0$). The analysis yielded that the tetravalent Ti ions occupy octahedral sites with distorted $O_h$ symmetry which is triggered by the presence of vacant 8*a* anionic site adjacent to $TiO_6$ octahedra. Further study with Fe doping revealed that it essentially reduces the octahedral distortion by introducing anionic disorder (migration of 48*f* oxygen ions to 8*a* site). Analysis of O *K* edge XAS spectra further confirmed the Fe substitution causing the systematic change in the ligand ($O^{2-}$) coordination of the $Ti^{4+}$ cations. On the other hand, a new field induced transition (with Fe doping) at low temperature $T^*$ (4 K<$T^*$< 8 K) in ac susceptibility with unusually slow spin relaxation was observed. The transition shifted towards higher temperatures both with increasing applied field and Fe concentration. However, the single ion spin freezing ($T_f$ ~35 K) appears to be suppressed with Fe substitution. Interestingly, small amount of $Fe^{3+}$ ion substitution showed significant enhancement in the dc magnetization at lower temperatures (<100 K). Analysis further indicated rise of dipolar FM exchange interaction with Fe doping.

## *I. Introduction*

The study of geometrically frustrated pyrochlore materials has received much attention due to its variety of low temperature ground states which is associated to the geometry of the spin lattice leading to frustration among the local spin interactions [1-3]. As a consequence of frustration, the spins cannot attain a long range order at low temperature but freeze in a random manner or remain dynamic down to lowest possible temperatures giving rise to exotic ground states. These special low temperature states include spin liquid states [4-9], apparent spin glass like states [10-15], spin ice state [16-21] and a state known as order by disorder [22-23] etc. These intriguing states are decided by balancing the strong crystal field effect (CFE), dipolar interactions and exchange interactions. The basic structure of pyrochlore $A_2B_2O_7$ is a derivative of fluorite type ($CaF_2$) structure which has two cations but one eighth fewer anions (i.e. an unoccupied interstitial site 8a surrounded by four $B^{4+}$ cations) [24-25]. The lattice parameter of the pyrochlore structure is double that of the fluorite structure. However, the pyrochlore structure is highly ordered as compared to fluorite where site disorder is present. It is often demonstrated as $A_2O^{'}.B_2O_6$ to express the interpenetrating network comprising of a cuprite-type tetrahedral net $A_2O^{'}$ along with corner-shared octahedra having composition of $B_2O_6$. An important parameter which decides the ordering of such $A_2B_2O_7$ structure is the co-ordination number. For, pyrochlore type structure, A (16d) and B (16c) ions are surrounded by 8 and 6 oxygen ions respectively while for fluorite type structure, B ions are prone to be in similar chemical surrounding like A ions (i.e. 8 $O^{2-}$ ions). In the ordered pyrochlore structure, the 48f oxygen ions are surrounded by 2A and 2B ions, 8b oxygen site is surrounded by 4A ions and 8a oxygen site is surrounded by 4B ions. As the ionic radii of A and B cations come closer, the structure gradually transforms into anion disordered fluorite structure by the migration of some oxygen ions from 48f (O1) and/or 8b(O2) sites to the vacant 8a sites. In the unit cell of pyrochlore $Eu_2Ti_2O_7$, the magnetic rare earth ions $Eu^{3+}$occupy 8 co-ordinate (16d) sites whereas the non-magnetic $Ti^{4+}$ ions occupy the six co-ordinate (16c) sites. Each $Eu^{3+}$ is surrounded by eight oxygen atoms, thus forming a trigonally distorted cube in which six O-atoms situated at 48f sites are lying on the equatorial plane of the cube while the rest two O-atoms (positioned at 8b) are lying diametrically opposite to each other along the <111> ($D_{3D}$) axis. On the other hand, the $Ti^{4+}$ ions (positioned at 16c) are lying adjacent to the vacant anion sites 8a and forming a distorted octahedron by the oxygen ions situated at 48f sites [24-25]. Obviously, the magnetic properties of these materials are mainly manifested by the electronic configurations of the *f* rare earth ions $R^{3+}$. Due to strong CFE, the ionic magnetic susceptibility of $R^{3+}$ ions differ along the $D_{3D}$ axis and along the axis perpendicular to it, which eventually raises the single ion anisotropy (SIA)[26]. In spin ice materials which follow "ice rules", the rare earth spins are highly uniaxial along <111> direction due to strong crystal field (CF) splitting driven single ion anisotropy(SIA). As a consequence, a highly degenerate macroscopic ground state is observed at very low temperature (<4 K) for all spin ice materials. This happens because any state that follows "two in/two out" ice rule for all the tetrahedra is a ground state which leads to a huge number of degenerate low energy states [20-21,28].Under this condition, the system cannot reach an ordered state by minimizing the dipolar interactions alone and hence ends up in a non-collinear, disordered and frozen state at very low temperature $T<T_{ice}$~4 K. However, it is relevant here to mention that, of all the spin ice materials $Ho_2Ti_2O_7$, $Dy_2Ti_2O_7$, $Ho_2Sn_2O_7$, $Dy_2Sn_2O_7$ etc.; only $Dy_2Ti_2O_7$ shows a higher temperature spin freezing at 15K along with its low temperature spin ice freezing at 4 K. This special higher temperature (~ 15 K) spin freezing for $Dy_2Ti_2O_7$ makes it more interesting towards various research communities and invited intense research attention towards this particular pyrochlore $Dy_2Ti_2O_7$. This 15 K spin freezing is attributed to single ion freezing and it is inter-linked with the low temperature ice freezing (<4 K) by a quantum tunnelling crossover process (which is characterized by a very weak temperature dependence of spin relaxation times thus showing a plateau region below 12 K) through the CF barrier and this is explained by creation and propagation of monopoles[29,30].

However, in our previous report on frustrated pyrochlore $Eu_2Ti_2O_7$, we have reported a new prominent spin freezing at around $T_f$ ~35 K, a temperature which is highest reported so far for such spin freezing due to geometrical spin frustration [31]. It has been inferred that the observed spin freezing is purely caused by local geometrical spin frustration where the spins have crystal field driven single ion anisotropy (SIA) which offers them less freedom for movement. The observed spin freezing is attributed to single spin relaxation process rather than a conventional spin glass freezing.

In the present pyrochlore material, the substitution of Fe (3*d*) ions in the rare-earth Eu (4*f*) site creates a platform for the different *f-f*, *f-d* and *d-d* interactions which inevitably add additional complexity in the system, thus making the magnetic properties more interesting. These 4*f*-3*d* interactions are reported to trigger several interesting phenomena in different systems including multiple magnetic transitions, spin reorientation transition; magneto-electric effect, magnetic field induced meta-magnetic transition etc [32-35]. It is pertinent to mention here that the configuration of 4*f* orbital is highly localized and much more complex than that of 3*d* orbital, thus the additional competing magnetic interactions introduced by the Fe (3*d*) spins (i.e*f-d* and *d-d*) is expected to drive the system in much more complex and interesting magnetic phases. In this context, we have incorporated Fe in ETO and investigatedthe change in dc and ac magnetic properties with Fe substitution in $Eu_{2-x}Fe_xTi_2O_7$ (EFTO). DC magnetic property is enhanced appreciably with increased Fe concentration. In ac susceptibility, the spin freezing around 35 K, appears to be suppressed with Fe substitution. Most interestingly, an additional clear peak associated to $Fe^{3+}$ ions is observed in ac susceptibility ($\chi^{//}$) study in presence of magnetic field at lower temperatures ($T^*$<15 K). The observed new $\chi^{//}$ peaks are purely field induced and associated to Fe doping. The spin relaxation responsible for the observed peaks is unusually slow. In the present paper, we have mainly focussed on investigating the field induced ac $\chi^{//}$ peaks by varying both the fields and compositions to probe the possible origin of the peaks.

Moreover, a prior knowledge to the electronic states of the constituent ions essentially helps towards the understanding of origin of the structural and magnetic properties, hence, we have studied the electronic structure of our samples by x-ray absorption spectroscopy (XAS) as well as x-ray magnetic circular dichroism (XMCD) measurements. The synchrotron based XAS is a spectroscopic tool to probe electronic states of a matter. We have performed XAS and XMCD measurements on this system as these are rarely reported in the previous literatures.

## *II. Experimental Detail:*

Polycrystalline $Eu_{2-x}Fe_xTi_2O_7$(with x=0.0, x=0.1 and 0.2) samples were synthesized using conventional solid state reaction techniques. The appropriate stoichiometric mixture of starting materials (high purity >99.99%) $Eu_2O_3$, $Fe_2O_3$ and $TiO_2$ were ground for half an hour and then heated at $1000^0$ C in air for 24 hours.The resulting powder thus obtained was reground and pressed into pellets. These pellets were re-heated at $1300^0$ C in air for several days with intermittent grindings and heating steps. X-ray powder diffraction (XRD) measurements confirm the prepared samples are of single phase with no chemical phase impurity. XRD measurements were performed using RigakuMiniflex II X-ray diffractometer. All the dc and ac magnetization studies were measured using a Quantum Design magnetic property measurement system (MPMS) super conducting quantum interference devices (SQUID) magnetometer. The XAS and XMCD measurements at Fe$L_{2-3}$ and Eu$M_{4-5}$ edges were carried out at the BL-14 beamline of Hiroshima Synchrotron Radiation Center (HiSOR), Hiroshima University, Japan. The XAS measurements at Ti$L_{2-3}$ and O $K$ edges were performed at Raja Ramanna Centre for Advanced Technology (RRCAT), Indore in INDUS 2 BL-1 beamline. In our case, total electron yield (TEY) mode has been used for recording XAS spectra because of its relatively easy experimental setup and high signal to noise ratio as compared to fluorescence yield.

## III. Results and Discussion

### A. *Structural study*

We examine the crystalline structure of the samples by refining the X-ray diffraction (XRD) data by Rietveld method using Fullprof software. The XRD pattern of all the three EFTO (x=0, 0.1, 0.2) samples along with its Rietveld refinement is shown in Fig. 1. It suggests that the samples crystallize in pure cubic phase (space group *Fd-3m* with the positions of Eu,Fe,Ti, O1 and O2 at 16d,16d, 16c, 48f, 8c respectively) without any traces of impurities. The crystallographic information obtained from refinement of XRD pattern is summarized in the Table 1. The values obtained for the goodness of fit (GOF) (which is defined by $R_{wp}/R_{exp}$, for a good fitting, the value should be less than 2) indicate the fitting to be reasonably good for all the samples. A pictorial representation of the crystal structure obtained through its refinement is shown in the upper inset of Fig. 1. The Rietveld refinement also shows linear decrease in lattice constants (a=b=c) with increasing Fe concentration as predicted by Vegard's law [36], thus suggesting a homogenous solid solution of Eu/Fe mixture [inset (bottom) of Fig.1]. This decrease in lattice parameter occurs because of the smaller ionic radius of $Fe^{3+}$ as compared to that of $Eu^{3+}$. As a result of Fe doping, the effective cell volume of the lattice as well as the effective Eu-Eu, Eu-O, Eu-Ti bond lengths decrease systematically and eventually modifies the exchange interactions which is manifested in their magnetic behaviours. Moreover, from the XRD analysis, the octahedral distortion is evident from the obtained bond angles measured on the equatorial plane of the octahedron. The estimated bond angles of O1-Ti-O1 are found to be $94.514^0$ and $85.486^0$ (which are observed to remain almost unchanged with Fe doping : Table-1) which are close to the bond angle values ~92.5$^0$ and ~87.5$^0$ reported for a similar pyrochlore with $TiO_6$ octahedral distortion [37]. However, the angle Ti-O1-Ti for pure system (x=0.0) ~132.97$^0$ has been found to decrease to ~132.38$^0$ with Fe substitution. This seems to be associated to the smaller size of the $Fe^{3+}$ ions (related to $Eu^{3+}$ ions) which in turn introduces additional strain in the system thus resulting in a slight change in the angle. The large distortion found in the $TiO_6$octahedra can presumably be attributed to the large difference in the ionic radii between the $Eu^{3+}$ and $Ti^{4+}$ ions.

### *B. X-ray absorption spectroscopy (XAS) study*

In order to probe the electronic structure and get insight into the local structural order of the present systems, X-ray absorption spectra have been recorded at room temperature. Figure 2(a) shows Fe 2*p* XAS spectrum corresponding to the photo-absorption from Fe2*p* core level to the Fe 3*d*or 4s unoccupied states (though 3d states are dominating) for 5% Fe doped sample (ETOF5). The Fe 2*p* XAS spectrum consists of two peaks Fe$L_3$(2$p_{3/2}$) and Fe$L_2$(2$p_{1/2}$) at ~710 eV and~724 eV, respectively, which is separated due tothe spin-orbit interaction ($\Delta$E~ 14 eV). It should be noted that small but distinct crystal field splitting in the $t_{2g}$ peaks of both Fe$L_3$ and Fe$L_2$ are discernible which indicates localized nature of Fe 3*d* electrons. The spectral features are essentially similar to the Fe2*p* XAS spectra of the well-studied $\gamma$-$Fe_2O_3$ system, where the Fe ions lie in tetrahedral and octahedral coordination with oxygen ions [38]. The shape of the XAS spectra is significantly different from that of $\alpha$-$Fe_2O_3$ where the splitting of Fe2p$3_{/2}$ is more pronounced due to the crystal field splitting as $Fe^{3+}$ ions occupy only the octahedral ($O_h$) sites [38]. Thus, it confirms the nominal valency of 3+ for Fe ions in the present system and also the successful substitution of Eu ions by Fe ions [38]. The Fe2*p* XAS spectral feature excludes contributions from the Fe 2+ spectral features as typically seen in metallic Fe, FeO or $Fe_3O_4$, suggesting absence of any mixed-valence states. The inset of Fig. 2(a) shows the XMCD spectra at Fe$L_{2,3}$ absorption edge which is obtained by taking difference between XAS spectra under +1T and -1T magnetic fields. We could not observe the XMCD signals at room temperature, indicating absence of magnetic ordering at room temperature.

Fig. 2(b) shows the Eu3*d* XAS spectrum at Eu$M_{4,5}$ edges at 300 K for ETOF5, which corresponds to the dipole transition from Eu3*d* core level to unoccupied Eu4*f* state. Two prominent peaks are observed corresponding to the Eu$M_5$ (3$d_{5/2}$→4*f*) at ~ 1131.2 eV and Eu$M_4$ (3$d_{3/2}$→4*f*)at~ 1158.9 eV, which are separated by the spin-orbit interaction (energy $\Delta$E~ 27.7eV). The Eu3*d* XAS spectra is similar to that of $Eu_2O_3$ [39], but different from that of EuO, confirming nominal valency of 3+ for Eu ions in our sample. The 3+ oxidation state of Eu ions can be evident from two clearly visible shoulder peaks A and B at ~1125.7 eV and ~1135.1 eV which are the signatures of 3+ oxidation states. Note that while $Eu^{3+}$ is non-magnetic in its $^7F_0$ ground state, its subsequent excited states $^7F_{1,2,3}$ are magnetic [26,39], leading to appreciable magnetic moment for this ion.

Fig. 2(c) depicts the room temperature Ti2*p* XAS spectra at Ti$L_{2,3}$ edges for all three samples (ETO [x=0.0], ETOF5 [0.1] and ETOF10 [0.2]). The Ti2*p* XAS spectrum is associated to the transition of electrons from Ti2*p* to Ti3*d* states. By comparing the line shape and energy separation between the obtained peaks of Ti$2p$ XAS spectra in $Eu_{2-x}Fe_xTi_2O_7$ with previously reported Ti$2p$ XAS spectra for $TiO_2$, $Ti_2O_3$, $BaTiO_3$,$SrTiO_3$, the observed spectra for all three samples are found to accord with those of $TiO_2$, $SrTiO_3$, $BaTiO_3$ suggesting nominal oxidation state is $Ti^{4+}$ [38,40-43]. The four peak structure of the observed spectra is a common feature for all tetravalent Ti based compounds with $TiO_6$ coordination, e.g. $TiO_2$, $SrTiO_3$, and $BaTiO_3$. One can see prominent four peak structures: the spectral features denoted as A, B correspond to the Ti$L_3$($2p_{3/2}$) absorption while C, D to the Ti$L_2$($2p_{1/2}$) absorption.The single sharp nature of the peak A observed at Ti$L_3$ edge is indicating the octahedral co-ordination of the $Ti^{4+}$ ions where Ti$3dt_{2g}$ orbitals are not involved in the bond formation. The Ti$L_{2-3}$ spin orbit splitting peaks are arising due to the transitions to the final states $Ti^{4+}$ $[(2p_{3/2},_{1/2})^{-1}$ $3d^1]$-$O^{2-}[2p^6]$, here the resulting hole in the $2p_{3/2}$ or $2p_{1/2}$ states is denoted as $(2p_{3/2},_{1/2})^{-1}$. The spin-orbit splitting energy is found merely to be ~5.35 eV for all the three samples. These spectral features are further split due to the crystal field, corresponding to the $t_{2g}$ and $e_g$ sub-bands. The $t_{2g}$–$e_g$ crystal field splitting peaks occur from the transitions to the final states $Ti^{4+}$ $[(2p_{3/2})^{-1}3d(2t_{2g})^1]$-$O^{2-}$ $[2p^6]$ and $Ti^{4+}$ $[(2p_{3/2})^{-1}3d(3e_g)^1]$-$O^{2-}$ $[2p^6]$ respectively. There is noticeable change in the crystal field splitting energies (10 Dq) with the Fe doping as they are found to be ~ 2.4 eV, 2.2 eV and 2.1 eV for x=0.0, 0.1 and 0.2 respectively. The changes in the crystal field splitting energies are associated to the changes in the local environment of the Ti ions with Fe substitutions which essentially alters its electronic structure. Interestingly, Ti$L_3$-$e_g$ peak further splits into an asymmetric doublet denoted by **B** and **B**$^/$. Similar $L_3$-$e_g$ splitting was observed in $TiO_2$, and it was attributed to the distortion in $TiO_6$ octahedra, which in turn gives rise to non-cubic ligand field effect [40-41]. However, $e_g$ states, consisting of two orbitals $d_{x^2-y^2}$and$d_{z^2}$, are very sensitive to any distortion from $TiO_6$ octahedra pointing toward the nearest neighbour oxygen atoms. It is relevant to mention here that the $L_3$-$e_g$ splitting into $d_{x^2-y^2}$ and $d_{z^2}$doublet for present system is similar to that of $TiO_2$, unlike $SrTiO_3$ or $BaTiO_3$ in which Ti is in perfect $O_h$ symmetry [38,40-43]. Interestingly, the octahedral distortion in the system was predicted from the XRD analysis which has been already discussed. Thus, the XAS data is supporting the previous XRD data analysis. Hence, this effectively elucidates that Ti has distorted $O_h$ symmetry in the present pyrochlore, similar to $TiO_2$. Eventually, the octahedral distortion is induced by the mismatch between the ionic radii of $Eu^{3+}/Ti^{4+}$ ions and due to the presence of oxygen vacancy at 8a site on the ab-plane adjacent to the $TiO_6$ octahedra [44]. In fact, the XAS spectra are more sensitive to the tetragonal distortion as compared to the trigonal distortion of the $TiO_6$octahedra [45]. It has been experimentally verified by the fact that for $TiO_2$ with rutile and anatase phases having tetragonal distortions of the $TiO_6$ octahedra, a clear asymmetric doublet is observed whereas a single symmetric peak is observed for ilmenite phase having trigonal octahedral distortion [45]. Thus, observation of the $e_g$ splitting in the present pyrochlore system suggests towards tetragonal distortion of the $TiO_6$ octahedra. It is interesting to note that the $e_g$ splitting is reduced with increased Fe doping ($L_3$-$e_g$ splitting of energies ΔE ~ 0.93 eV, 0.72 eV and 0.63 eV are found for samples x=0.0, 0.1 and 0.2 respectively). This suggests that octahedral distortion in $TiO_6$ octahedra diminishes by Fe substitution while the Ti ions still remain in octahedral symmetry. Moreover, another feature which can be observed (Fig. 2c) is that the relative intensities between the $t_{2g}$ and $e_g$ peaks are getting changed systematically with increased Fe substitution. It is seemingly associated to the increased overlapping of transitions to the $e_g$: $d_{x^2-y^2}$ and $d_{z^2}$ states. As a matter of fact, for ordered pyrochlore structure, the difference in ionic radii between the A and B site ions should be large otherwise it transforms in to the disordered fluorite structure (if their ionic radii become similar)[46]. Hence, as $Fe^{3+}$ ions are substituted in the $Eu^{3+}$ site, due to the similar ionic radii of $Ti^{4+}$ and $Fe^{3+}$, it is expected to trigger the local anti-site anionic disorder due to the changes in the local oxygen (mainly 48*f* site oxygen is responsible) site in the system [44,46]. Factually, if we look at the Ti$L_2$ edge, the peak position of peak D is observed to shift towards lower energy i.e. energy separation ΔE between C and D peaks is decreased with Fe substitution. Again, the intensity height ratio of these two peaks C and D (C peak is fitted with single peak with intensity $I_C$ while D peak is fitted with two peaks with intensities $I_{D1}$ and $I_{D2}$ due to its shoulder at higher energy i.e. the ratio is $I_C/I_{D1}+I_{D2}$) is found to diminish with Fe doping (not shown). Similar findings were observed in a Zr based pyrochlore system $Nd_{2-x}Y_xZr_2O_7$ due to substitution of smaller cations $Y^{3+}$ in the $Nd^{3+}$ site [46].Where, the local anionic disorder was gradually introduced due to site change of 48*f*/8*b* oxygen ions to 8*a* sites which eventually increased the local co-ordination of Zr ions from 6 to 7 (probed by X-ray absorption spectroscopy)[46]. Thus, the observation of such decreasing trend in the peak splitting energy and the intensity ratio with Fe substitution in the present $Eu_{2-x}$FexTi$_2O_7$ can be presumably attributed to the local anionic disorder caused by the 48*f* site oxygen migration towards the 8*a* site which effectively altered the local environment of $Ti^{4+}$ ions. The change in the local environment of the $Ti^{4+}$ ions is also evident in consequent changes in the $L_3$-$e_g$ splitting where the splitting is diminished with Fe doping, suggesting a decrease in the octahedral distortion in the $TiO_6$ octahedra. As Fe is doped in Eu site, due to similar ionic radii of the $Ti^{4+}$ and $Fe^{3+}$, the octahedral distortion in the $TiO_6$ octahedra is expected to be diminished. Thus it is plausible to state that as $Fe^{3+}$ ions are substituted, due to the migration of oxygen ions (positioned at 48*f* or 8*b*) to the vacant 8*a* sites (which is triggered by the comparable ionic radii of $Fe^{3+}$ and $Ti^{4+}$), the octahedral distortion is reduced and the anionic disorder is introduced [44,46]. However, the observation of relatively less $L_3$-$e_g$ splitting for present systems as compared to that of $TiO_2$ (ΔE~ 0.93 eV) suggests for relatively a smaller distortion from the perfect $O_h$ symmetry for all the three samples [40].

However, this explanation of $e_g$ splitting still remains under debate because the distortion effect is too weak to produce significant $e_g$ splitting [47]. Later, P. Kruger claimed (by first-principles calculations on large number of clusters of $TiO_6$) that the $e_g$ splitting is not due to the local structure effects, but rather arises due to a long range crystal structure effect[48] and showed that experimentally observed $L_3$-$e_g$ splitting could only be reproduced by taking at least an array of 60 atoms and hence this is a "non-local" effect having a length scale of 1 nm. Thus, it may also be plausible to explain the observed diminishing feature of the $L_3$-$e_g$ splitting (with increased Fe doping) as increased Fe doping introduces more disorder (due to random $Fe^{3+}$ ions substitution at A site and partial migration of some oxygen ions to 8*a* sites) in the system, long range structural ordering gets affected which changes the $L_3$-$e_g$ patterns.

Two weak satellite peaks marked as S1and S2 exist at photon energies ~473 eV and ~476.7 eV can be found in Ti$L_{2,3}$ XAS spectra of our systems. Similar satellite peaks were reported in Ti2*p* XAS spectra of $TiO_2$. These satellite peaks can be presumably attributed to the effect of hybridization of Ti 4*p*, Ti 3*d* and Ti 4*s* states with ligand orbitals O 2*p*. M. Umeda *et al.* explained the origin of these types of satellite peaks through charge transfer (CT) process between O 2*p* valance band and unoccupied 3*d* states of transition metal ions to form hybridized intermediate states [49]. Eventually, the intensity and the position of these satellite peaks are decided by the hybridization energy between transition 3*d* and ligand O 2*p* states. It has been reported that the peak positions and the relative intensities of these satellite peaks can be theoretically reproduced using charge transfer energy and hybridization energy [50].

Fig. 2(d) shows the room temperature XAS spectra recorded at O *K* edge for all the three samples $Eu_{2-x}FexTi_2O_7$ (x=0.0, 0.1 and 0.2). The spectral feature of O K edge spectra consists of two main regions:1) the region between the energy ~530 eV to 536 eV that arises due to the hybridization of empty O*2p* orbitals to the Ti*3d*, Fe*3d* and Eu*4d* orbitals and this region is very sensitive to the ligand co-ordination and the local crystal symmetry. 2) the region above ~ 537 eV results from the longer overlap of O*2p* orbitals to the Ti, Fe 4*sp* and Eu*5sp* orbitals, is associated to the long range structural ordering of the system [41,44]. The two peaks A and B in the first region are identified as $t_{2g}$ and $e_g$ states which arise from the transitions to the final states $Ti^{4+}/Fe^{3+}$ $[3d(2t_{2g})^1]$-$O^{2-}$ $[(1s)^{-1}2p^6]$and $Ti^{4+}/Fe^{3+}$ $[3d(3e_g)^1]$-$O^{2-}$ $[(1s)^{-1}2p^6]$, where $(1s)^{-1}$ is denoting a hole created at the O 1s shell. However, the O K edge spectra correspond to the transition from O 1s core level to the unoccupied O *2p* states hybridized with the empty metallic d (mostly) and s-p states following the dipole selection rule. Thus the observation of the intense A and B peaks suggests that all the associated metallic bonds Ti-O, Fe-O are of covalent in nature in which unoccupied density of states exists in hybridized O*2p*-Ti(Fe)*3d* states. On the other hand, it is comprehensible that if the bonds were of ionic in nature, no such absorption peaks were found since they correspond to fully occupied O*2p* orbitals. It is interesting to note that the energy of splitting (ΔE→ crystal field splitting) between $t_{2g}$ and $e_g$ (A and B) peaks decreases gradually with increasing Fe substitution. Here, the position of the A peak is observed to remain merely fixed while the B peak shifts towards the lower energy (535.1 eV to 534.5 eV) which is consistent with the Ti$L_{2-3}$ edge spectra. Again, this effect can be attributed to the change in the local environment (increase in the co-ordination) around $Ti^{4+}$ ions which is triggered by the anionic disorder introduced by the substitution of Fe ions with lower ionic radii [46]. Apart from this, the change in the splitting energy is expected due to the increasing overlap of the Fe*3d* states with the existing Ti*3d*-O*2p* hybridized states which is evident in the observed spectra. Moreover, looking at the second region (>537 eV), we can observe a prominent peak C which is followed by two broad features D and E which are originated from the hybridization of O*2p* and metallic *s-p* states [41]. Again the shift in the peak position observed for the peak C can presumably be ascribed to the additional contribution coming from Fe*4s-p* states to the existing Ti*4s-p* and Eu*5s-p* states. A noticeable change can be observed in the broad features D and E which are flattened and gradually become weak as Fe concentration is increased. This is a clear indication of increasing disorder in the system (as broadening of these spectral features are known to be related to the diminishing long range structural order) which is being introduced by random Fe substitution in the Eu site [51].

### *C. DC magnetization study*

DC magnetization measurements of the samples $Eu_{2-x}FexTi_2O_7$ (x=0, 0.1, 0.2) were carried out as a function of temperature and magnetic field. Fig. 3 (a-c) shows the temperature dependence of magnetization (M) following the zero field cooling (ZFC) and field cooling (FC) protocols with applied magnetic field of 100 Oe for all the three samples. In M(T) curve for pure sample (x=0.0), (Figure 3(a)) it is observed that M increases with decreasing temperature from 300K and reaches a maximum at around~90 K. As the temperature decreases further, a slope change can be observed and the magnetization curve enters in a plateau region. This plateau extends down to ~ 20 K, below this temperature a sharp increase in magnetization is observed. The observed plateau can be ascribed to the strong crystal field effect [52]. For the whole temperature range (2-300 K), the M-T curve for pure sample i.e. x=0.0 does not follow the standard Curie-Weiss (CW) law: $= \frac{M}{H} = \frac{C}{T-\theta_{CW}}$ , here C is the Curie constant and $\theta_{CW}$ is the CW temperature. Eventually, an attempt to fit the M-T curve with Curie-Weiss law below 100K was successful, suggesting the system remains paramagnetic down to lowest available temperature ~2 K (Inset of Fig. 3(a)). However, the small bifurcation or thermo-magnetic irreversibility observed between the ZFC and FC curves suggests existence of spin frustration in this system. In the earlier report, it was seen that application of higher magnetic field ~1 T caused the two curves to merge completely, confirming the underlying spin frustration to be different than spin glass type [31]. The magnetization value is drastically enhanced with $Fe^{3+}$ ion substitution as evident in fig. 3 (b-d). As a consequence of $Fe^{3+}$ion substitution, the temperature independent plateau region is disappeared. Eventually, the magnetization increases with decreasing temperature and no magnetic phase transition is observed which can be ascribed to the positive *f-d* super exchange mechanism [53]. The enhancement of magnetization may presumably be attributed to the exchange interactions between Eu-Fe and Fe-Fe spins and to the larger moment of $Fe^{3+}$ spins. As a matter of fact, in each tetrahedron of the pyrochlore system, there is on an average very few $Fe^{3+}$ ions sitting on the lattice points since the concentration of $Fe^{3+}$ doping is very low (up to 10%). Thus, $Fe^{3+}$ ions cannot form its nearest neighbour pair with another $Fe^{3+}$ ion, as a consequence the direct exchange interaction between Fe-Fe is weak. Hence, Eu-Fe spin interactions (*f-d*) play the main role in altering the magnetic behaviour of the systems. Moreover, as evident from XRD analysis, the Fe substitution reduces the effective Eu-Eu distances which in turn enhance the exchange interactions among these spins. As can be seen from Fig. 3 (b –c), the Fe doping (x=0.1 and 0.2) results in complete coincidence of ZFC and FC curves with each other unlike the pure sample (x=0.0) which suggests that the Fe substitution relieves the spin frustration in the system. However, if we take a closer view in the enlarged section of the ZFC FC curves (as shown in the sub-inset of Fig. 3b and c) very small (relative to that for pure system) thermo-remnant irreversibility can still be found, suggesting the presence of weaker spin frustration in the doped systems i.e. x=0.1 and 0.2. However,the M(T) curves for both the Fe doped samples (x=0.1 and 0.20) were found to get well-fitted by Curie-Weiss law above ~ 100 K, this indicates that the system remains paramagnetic down to ~100K (inset of Fig. 3 (b and c)). However, as evident from inset of Fig. 3(b-c), below the temperature ~100 K, the experimental M(T) curves start deviating from standard Curie-Weiss behaviour. This directly suggests that the Fe doped system (x=0.1 and 0.2) does not remain simply paramagnetic instead ferromagnetic contribution comes into play below100 K. Factually, the sharp rise in magnetization below ~20 K is seemingly associated to some type of magnetic ordering. Hence, to confirm the nature of magnetic ordering, another attempt to fit the M(T) data for the Fe doped samples (x=0.1 and 0.2) in the lower temperature region (2-100 K) was made with standard three dimensional spin wave model: M(T)=$M(0)(1-AT^{\frac{3}{2}})$, where M(0) is the saturation magnetization related to the ferromagnetic contribution at T=0 K and A being a parameter associated to the structural properties of the system [54]. However, the fitting was not satisfactory for both the Fe doped samples i.e. x=0.1 and 0.2 (not shown). Interestingly, a combination of these two models i.e. M(T)=$\frac{(C\times H)}{T-\theta_{CW}}+M(0)(1-AT^{\frac{3}{2}})$ was found to result in a reasonably good fitting for both the samples as can be seen from the inset of Fig. 3(b-c).The first term of the above expression which is related to the paramagnetic contribution dominates at relatively higher temperatures (>20 K) whereas the second term pertaining to the ferromagnetic component dominates in the comparatively lower temperature region (<20 K) where sharp rise in magnetization occurs. Thus, the above fitting clearly suggests the existence of ferromagnetic contribution along with the paramagnetic phase at lower temperatures (<100 K). To further investigate how the magnetization in this (lower temperature) region (< 20 K) varies with magnetic fields, we have recorded M(T) curves with higher magnetic fields as shown in the inset 1 and 2 of Fig. 4 for x=0.1 and 0.2 samples respectively. From the figures, it is evident that the magnetization behaviour at lower temperature region shows very strong field dependence for both the Fe doped samples. Typically, such strong field dependence of the magnetization data also suggests towards the existence of magnetic ordering in the lower temperature region.

Understanding the above facts, the sharp rise in magnetization at low temperatures cannot be attributed to the CFE, hence other magnetic interactions, e.g., exchange interactions, dipolar interactions etc., should be taken into considerations to examine the origin behind such behaviour. Therefore, to calculate the contributions of these different magnetic interactions, high temperature series expansion of the susceptibility $\boldsymbol{(\chi) = C\,[\frac{1}{T} + \frac{\theta_{cw}}{T^2}]}$ was considered [52,55]. From linear fit of the plot "$\boldsymbol{\chi T}$ Vs $\boldsymbol{\frac{1}{T}}$" for the temperature range 2-5 K, the values of Curie Weiss temperature $\Theta_{cw}$, effective magnetic moment $\mu_{eff}$, exchange interaction energy $J_{nn}$ and dipolar interaction energy $D_{nn}$ were calculated (inset a,b and c of Fig.3 (d)). The exchange interaction energy was calculated using the relation $J_{nn} = \frac{3\theta_{cw}}{zS(S+1)}$ , where z=6 be the co-ordination number. $\mu_{eff}$ was determined using the relation $C= \frac{N\mu_{eff}^2}{3K}$ and the dipolar interaction energy $D_{nn}$ was obtained from $D_{nn}= \frac{\mu_{eff}^2\mu_0}{4\pi r_{nn}^3}$ where $r_{nn}$ refers to the distance between a $Eu^{3+}$ ion at (000) and its nearest neighbour at (*a*/4,*a*/4,0), *a* being the lattice constant of the unit cell [52,56]. The calculated values of these parameters ($J_{nn}$, $D_{nn}$, $\mu_{eff}$, $\Theta_{cw}$) determined from the aforementioned formulae are summarized in Table-2. The data obtained for pure (x=0.0) sample indicates nearest neighbour AFM exchange interaction ($J_{nn}$) dominating over ferromagnetic dipole-dipole interactions ($D_{nn}$) while the Curie-Weiss temperature is found to be slightly negative (-1.35 K). These results are in well agreement with the previous report [52]. Apart from this, it is clear from Table-2 that with increasing Fe doping, FM dipolar interaction and effective magnetic moment increase appreciably. Interestingly, as is observed in Table-1, Fe substitution though initially suppresses the AFM exchange interactions $J_{nn}$ for x=0.1 but with increasing Fe substitution x=0.2, $J_{nn}$ is not further diminished, instead it increases a bit. Similarly, the Curie-Weiss temperature $\Theta_{cw}$ also initially becomes less negative with x=0.1, then for x=0.2, it shows a little increment (almost constant) but for both the cases, $\Theta_{cw}$ remains less negative than that of x=0.0 sample. In the present scenario, the possible explanation of the above magnetic behaviour can be related to the Eu -Fe spin interactions. Initially for x=0.1 sample,$\Theta_{cw}$becomes less negative suggesting a rise in ferromagnetic interactions in the system. This induced ferromagnetism is possibly due to the interactions between $Eu^{3+}$ -$Fe^{3+}$ (as is clear from XAS analysis). In this case due to very small Fe concentration the AFM $Fe^{3+}$- $Fe^{3+}$ interactions remain very weak. However, as Fe concentration increases to x=0.2, the AFM $Fe^{3+}$- $Fe^{3+}$interactions slightly increase, thereby pushes the$\Theta_{cw}$towards slightly more negative side. However, it is observed that for x=0.2 the dipolar energy increases by two-fold to that of x=0.1 sample(as is observed in Table-1). This large dipolar energy in x=0.2 sample may also be responsible for the larger magnetic moment at lowest temperature of measurement.

To further investigate the nature of the magnetic ordering induced due to the Fe doping, the "isothermal field (H) variation of magnetization (M)" measurements are carried out. We have recorded the M-H curves at temperature 2K for all the $Eu_{2-x}FexTi_2O_7$ (x=0.0, 0.1, 0.2) samples as is shown in Fig. 4. The unsaturated linear nature of the M-H curve for the pure (x=0.0) sample suggests that the paramagnetic phase prevails even at such low temperature ~2 K. However, the non-linearity of M(H) curves (S-shaped) for Fe doped samples (x=0.1, 0.2) indicates effective rise of ferromagnetic spin ordering in the system. On the other hand, unsaturated nature of the magnetization upto 2 T, indicates strong antiferromagnetic interactions are also present in the system which was also predicted from our previous M(T)data. Again, as already observed from M(T), it is again evident from M(H) loops that with increasing Fe concentration the magnetization (M) also increases. A close observation reveals the coercivity of the hysteresis loop also increases with the Fe doping (inset 3 of Fig. 4). Hence, it supports the results obtained from dc susceptibility analysis that at low temperatures, ferromagnetic contribution dominates due to complex $Eu^{3+}$- $Eu^{3+}$, $Eu^{3+}$- $Fe^{3+}$ and $Fe^{3+}$- $Fe^{3+}$ interactions which consist of isotropic, anisotropic symmetric and anti-symmetric exchange interactions.

### *D. AC magnetization study*

In contrast to dc magnetization study, ac-susceptibility study with different frequencies and fields allows us to probe the spin relaxation process [57]. We have already reported a spin freezing transition for pure (x=0.0) sample, as a characteristic behaviour in its higher temperature ac susceptibility ($\chi^{/}$ and $\chi^{//}$),[inset of figure-5(c) and 5(d)][31]. The spin freezing was observed as a sudden drop in ac susceptibility $\chi^{/}$(in phase part), while its corresponding Kramers-Kronig frequency dependent peaks were found in ac $\chi^{//}$(quadrature part) below 35 K (for the measured frequencies). The frequency dependence of the $\chi^{//}$ peaks related to the spin relaxations occurring near this spin freezing temperature $T_f$~35 K, follows Arrhenius law $\boldsymbol{f = f_0 e^{-E_b/K_B T}}$ , thus is a thermally activated process with an activation energy barrier $E_b$=339 K. Single spin relaxation process was found to be responsible for the observed spin freezing. Since, the pyrochlore materials are highly ordered having minimal structural disorder or site randomness (<1%), (while the structural disorder is believed to be the origin of conventional spin glass transitions) observation of this new type of spin freezing other than spin glass was not very surprising [58].

To examine how the spin relaxations of rare earth $Eu^{3+}$ spins get modified with $Fe^{3+}$ ions substitution, we have studied detailed ac susceptibility measurements for the doped samples $Eu_{2-x}FexTi_2O_7$ (x=0.1, 0.2). Interestingly, we observe, even small concentration of Fe substitution (5%) have greatly changed the spin relaxation by introducing $Eu^{3+}$-$Fe^{3+}$ magnetic ion interactions and altered crystalline fields. In Fig. 5 (a&c), we have shown the temperature variation of ac $\chi^{/}$ and $\chi^{//}$of Fe doped sample (x=0.1) without any applied DC magnetic field i.e. H=0 Oe. The substitution of as little as 5% Fe, has caused a dramatic change in the nature of the $\chi^{/}$ curves, as the frequency dependence of $\chi^{/}$curves have become very weak while its drop below the freezing temperatures ($T_f$ ~ 35 K) is almost disappeared (Fig. 5a). The curve $\chi^{/}$ (x=0.1) shows a monotonous rise with increasing slope as temperature decreases for all the excitation frequencies of our study, a feature which is completely different from the x=0.0 sample. This is in strong contrast to the magnetic behaviour observed in the non-magnetic diluted system where the spin freezing transition was observed to be enhanced [31]. Eventually, little substitution of $Fe^{3+}$ ions in $Eu^{3+}$ site causes local disorder which is sufficient to disrupt the crystalline field spacing of $Eu^{3+}$ ions, consequently it results in significant suppression of the observed spin freezing. The suppression of the spin freezing in $\chi^{/}$, clearly indicates a faster spin relaxation occurring in the Fe substituted system. A close observation at $\chi^{/}$ curve, shown in inset of Fig. 5(a), reveals that weak frequency dependence is still present in the system. Though signature of the spin freezing is very feeble in $\chi^{/}$ curves, the corresponding clear peaks of $\chi^{//}$ are present as shown in Fig. 5(c), indicating the presence of spin relaxation which is still thermally driven.

Fig. 6 shows the Arrhenius fit to the temperature dependent peaks $\chi^{//}$at different frequencies, which gives a thermal energy barrier $E_b$= 316K (x=0.1). The decrease in the energy $E_b$, possibly due to the little modification in the CF levels and change in the spin-phonon spectrum associated with small changes in lattice constant and electronic structure of the system [59]. The plausible explanation for the faster spin relaxation thus can be given, as the Fe doping reduces the thermal energy barrier, it takes less time for the spins to relax, thus we observe relatively faster relaxation. However, a close observation reveals a minute shift of the freezing temperature $T_f$ towards lower temperature with Fe doping (inset of Fig. 6).

To study the influence of an external dc field on the spin relaxation process, we have applied different dc magnetic fields (H) during ac susceptibility measurements. To our surprise, we have observed an additional clear dip producing a local maxima in $\chi^{/}$ curve appearing around a temperature $T^{*}$~ 4K, for H=1T while the corresponding peak in $\chi^{//}$has also appeared in a narrow and sharp shape [Fig 5(b,d)]. Clearly this new peak is associated to a dissipative spin relaxation process which is solely field induced andit is different from the previously reported single ion spin freezing observed at a higher temperature $T_f$(~35 K). The absence of frequency dependence in these field induced $\chi^{//}$(T) peaks clearly rules out the thermal relaxation process of the spins which follow Arrhenius law. Since, no such field induced transition is observed at low temperature in pure pyrochlore $Eu_2Ti_2O_7$; the transition must solely be associated with $Fe^{3+}$ ion. As a matter of fact, site-randomness is believed to be a source of typical spin glass freezing and in this case, the Fe substitution may cause random occupancy of Eu and Fe spins which in turn can give rise to spin glass type freezing. However,the newly observed $\chi^{//}$ peaks do not appear as a glassy transition, firstly as the typical spin glass peaks are quite broad and they are frequency dependent while the observed $\chi^{//}$ peaks are sharp and merely frequency independent (Fig.5d). Secondly, the temperature variation of dc magnetization curves (for x=0.1 sample) has neither showed any anomaly related to spin glass transitions nor any bifurcation was observed between ZFC and FC curves near this temperature $T^{*}$. Hence, the origin of the observed transition seems to be something else than spin glass. However, to further elucidate the intriguing physics regarding the observed peak, we have investigated the ac susceptibility measurements with increasing fields and increased Fe concentration (x=0.2). It is visible in the Fig. 7(c), with increasing field H, the freezing temperature gets shifted towards higher temperature, which is in a strong contrast to a spin glass transition [60]. This further cancels out the possibility of getting spin glass transition in the observed peaks. It can be noted that with increasing field H, the drop in $\chi^{/}$ increases progressively forming a local maxima below 10K (Fig. 7a).

To investigate the effect of increased doping (Fe) concentration, we have performed similar ac susceptibility measurements on x=0.2 sample. In Fig. 8(a-d), the variation of $\chi^{/}$(T) and $\chi^{//}$(T) curves for x=0.2, are shown at different frequencies and fields. At H=0 Oe, the behaviour of $\chi^{/}$ and $\chi^{//}$ curves appears to be similar to the x=0.1 sample, with similar decreased drop in $\chi^{/}$ and very weak frequency dependence is observed, [Fig. 8(a-c)]. However, for this case also, spin relaxation is still present near characteristic freezing temperature $T_f$ ~35 K, which is evident from $\chi^{//}$ curves [Fig. 8(c)]. Interestingly, for x=0.2, the thermal energy barrier $E_b$ as extracted from the Arrhenius fit to the frequency dependence of $\chi^{//}$ (Fig. 6), has a smaller value $E_b$=304 K. This eventually suggests a monotonous decreasing trend in CF level spacing with increasing Fe substitution. Again for this sample also, it can be noted that the freezing temperature $T_f$ decreases further a bit (inset of Fig. 6). To further investigate whether the fundamental nature of this spin freezing transition gets modified with Fe doping, we have used Mydosh parameter (p) which is typically used for confirming a spin glass transition: $= \frac{\Delta\tau_f}{\tau_f \Delta log f}$; where $\tau_f$ is the freezing temperature for the frequency *f* [31]. The earlier reported value of *p* for the pure sample i.e. x=0.0 was ~0.286. Whereas for both the doped samples (x=0.1 and x=0.2) the parameter *p* has been reduced to ~0.2. On the other hand, for conventional spin glass transition the value of *p* should be of the order of ~0.01 which is a much smaller value than the *p* values obtained for all of our three samples (x=0.0, 0.1 and 0.2). Thus, the observed spin freezing transition is not of typical spin glass type. Hence, it can be inferred that the fundamental nature of the single ion spin freezing near ~35 K remains unaffected even after 10% Fe substitution. The little decrease in the Mydosh parameter *p* can be understood on the basis of the random occupancy of the doped Fe ions which usually contribute towards the typical spin glass state.

However, on application of dc fields H, similar local maxima in $\chi^{/}$ and the associated peaks in $\chi^{//}$($T^*$ ~6 K to 8 K) starts appearing as observed in previous case (x=0.1), [Fig. 8(b&d)]. The only noticeable difference that can be noted that for x=0.1, the peak position of field induced $\chi^{//}$ peak is at~ 4 K (H=1 T) and for x=0.2, the position is at ~ 6 K (H=1T). Hence, Fe substitution causes the $\chi^{//}$ peak positions to shift towards higher temperature. Surprisingly, though the field induced transition shifts its position with Fe doping, the single ion spin freezing peak is observed to remain unshifted. It should be mentioned here, in $Dy_xTb_{2-x}Ti_2O_7$ (DTTO) hybrid system, two distinct field induced peaks in ac susceptibility were reported, where higher temperature peak is around 16K (associated to single ion freezing) and a new peak near 12K (varies with applied dc field and Dy concentration) [61]. The origin of the field induced peak (~12 K) which is associated to $Tb^{3+}$ is still not clear, as it showed unusual jump in the thermal energy barrier with small amount of $Tb^{3+}$ substitution. There are qualitative differences between field induced peaks reported in DTTO and peaks observed in the present $Eu_{2-x}FexTi_2O_7$ (EFTO). Both the observed peaks in DTTO are thermally activated and they are emerged when external field is applied. Whereas for the EFTO, the higher temperature peaks ($T_f$ ~35 K) are thermally activated but the lower temperature (4 K<$T^*$<8 K)) field induced peaks don't follow Arrhenius behaviour and thus they are not thermally activated, hence arenot related to CF states of $Eu^{3+}$. Again, though the lower temperature (4 K<$T^*$<8 K) peaks are originated by external field influence, the higher temperature ($T_f$~35 K) peaks are not enhanced by external fields unlike in DTTO. Most interestingly, one can note that unlike the higher temperature ($T_f$~35 K) freezing in EFTO, where upon decreasing frequency the ac $\chi''$ peaks get suppressed, the lower temperature field induced $\chi''$ peaks show monotonous growth even at 10 Hz, (Fig. 7(d)). The growing $\chi''$ peaks at such a low frequency (10 Hz) clearly suggests an unusually slow relaxation of spins at this low temperature region (4 K<$T^*$<8 K). It is in strong contrast with common convention that strong applied magnetic field would cause a faster spin relaxation to achieve an equilibrium spin state. However, the observed phenomena could be explained in different ways. To elucidate the observed slow relaxation, the field induced local maxima observed in $\chi'$ may be addressed to "single moment saturation" which states "in presence of a sufficiently high field, the susceptibility associated to thermal fluctuations of spins approaches zero both at high temperatures and as T→0 K" [62]. Such field induced "single moment saturation peaks in $\chi'$ " were previously reported in spin liquid $Tb_2Ti_2O_7$(TTO) and spin ice $Dy_2Ti_2O_7$ at comparatively higher temperatures(T>20 K) [62]. The similar kind of unusually slow spin relaxation was observed in these systems. The reason behind such slow relaxation can be explained by the formation of correlated regions inter-linked by strong dipolar interactions in these strongly polarized systems, where some spins are polarized along and some are oppositely polarized to the applied field. Hence due to the presence of large energy barrier to switch these oppositely oriented spins, the relaxation becomesextremely slow, thus it can even be detected by ac susceptibility even at such a low frequency of 10 Hz. Earlier, in dc magnetization study, we have observed rise of dipolar interaction with Fe substitution (Table 1), hence it can in turn give rise to such correlated spin polarized regions, which may thus raise such unusually slow spin relaxation. Thus the single moment saturation effect can elucidate this low temperature field induced transition quite reasonably. However, the dipolar interaction is still not strong enough as compared to other strongly correlated paramagnets like TTO. However, on looking at the narrow sharp distribution of these $\chi''$ peaks, an alternative explanation which can be given is that it may indicate some type of magnetic ordering at low temperature which was manifested in sharp rise in dc magnetization curves at low temperatures. It is relevant to note here that rise of ferromagnetic ordering was already suggested by the dc M(T) data analysis by spin wave model. As mentioned before, the increasing applied magnetic field causes the peaks $\chi^{//}$(T) to shift towards higher temperature indicating rise of ferromagnetic interactions[63]. Again minimal frequency dependence of the $\chi^{//}$(T) peaks also suggest a magnetic ordering. Hence, upon inspection on the sharp nature and the frequency independence of these field induced $\chi^{//}$ peak positions, development of ferromagnetic ordering can be a possible origin of the observed transition. Another rare possibility of getting slow spin relaxation may be due to the change in spin-phonon interactions through Fe substitution. Since the mass difference between the two elements is quite large i.e. Eu is almost 3 times heavier than Fe, the Fe substitution would cause appreciable change in the spin phonon spectrum of the system which can result in slow relaxation of the spins.

***Conclusion***

We have synthesized and characterized the pyrochlore compound of $Eu_{2-x}Fe_xTi_2O_7$ with partial substitution of Fe on Eu site. The compounds adopted a cubic pyrochlore structure with *Fd-3m* symmetry and the linearly decreasing trend in lattice parameter with Fe substitution suggests homogeneity of the sample. Electronic structure of EFTO has been investigated using XAS measurements. The analysis confirmed the elemental nominal oxidation statesto be $Eu^{3+}$, $Fe^{3+}$ and $Ti^{4+}$. Clear splitting of $L_3$-$e_g$peak of Ti 2p XAS spectra can be attributed to the deviation from pure $O_h$ symmetry in Ti octahedral coordination in EFTO which is similar to the octahedral distortion present in the rutile and anatase $TiO_2$ phase. The octahedral distortion was also evident from the X-ray diffraction analysis as it showed sizeable distortion present in the $TiO_6$ octahedra. However, the octahedral distortion is originated from the absence of interstitial anionic vacant site 8*a* lying close to the $TiO_6$ octahedra. Further investigations showed that the octahedral distortion gets diminished appreciably with the Fe doping which is attributed to the anionic disorder introduced by the comparable ionic radii of $Ti^{4+}$ and $Fe^{3+}$ (which is required for forming disordered fluorite structure). Both the Ti$L_{2-3}$ and O *K* edge XAS spectra analysis showed that the effective co-ordination of the $Ti^{4+}$ ions gets altered with Fe doping which is seemingly associated to the migration of 48*f* and 8*b* oxygen migration to the vacant 8 site (adjacent to $TiO_6$) leading to the anionic disorder. This anionic disorder in turn reduces the octahedral distortion of the $TiO_6$octahedra. In the thermo-magnetometry study, dc magnetization data showed an indication towards enhanced ferromagnetic interactions on Fe doping which is mainly due to the Eu-Fe interactions. The Fe doped systems (x=0.1, 0.2) remained paramagnetic down to ~100 K following the Curie-Weiss behaviour (>100 K). However, sharp deviation from Curie-Weiss law below ~100 K, suggested the onset of magnetic ordering below this temperature. Presence of ferromagnetic contribution below ~100 K is confirmed from the combination of three dimensional spin wave model and Curie-Weiss law. Ac susceptibility study showed clear suppression of single ion freezing ($T_f$ ~35 K) with Fe doping, however, the faster spin relaxation related to $Fe^{3+}$ ions is responsible for the observed freezing suppression. A small but systematic decrease in freezing temperature $T_f$ is observed with Fe substitution. However the monotonous decrease in the thermal energy barrier $E_b$ of spin flipping suggests modification in the local environment of $Eu^{3+}$ spins which alters the crystalline fields of $Eu^{3+}$ spins. Despite of the suppression in the single ion freezing (~35 K) peaks, the fundamental nature of this spin freezing remained unchanged which is different than spin glass state. Most interestingly, ac susceptibility with non zero magnetic fields revealed a prominent $\chi''$ peak (which is purely associated to $Fe^{3+}$ ions) at lower temperatures (4 K<$T^*$<8 K) in addition to the single spin freezing peak (at $T_f$~35 K). These $Fe^{3+}$ related peaks are not thermally driven and possibly not related to the crystalline field unlike single ion peak at $T_f$~35 K. This new field induced peaks are frequency independent and relatively quite sharp unlike the single spin freezing peaks. The peak positions are observed to shift towards higher temperature both with Fe doping and increasing fields. The field induced peak in $\chi''$ is not caused by typical spin glass freezing. Regardless of the underlying physical origin, the demonstrated sudden suppression of the single spin freezing peak $T_f$ and emergence of field induced peak $T^*$ through Fe substitution suggest that the suppression arises from proper balance of different Eu-Fe (*f*-*d*) interactions in the system. Detailed frequency and field dependent ac susceptibility study showed the associated spin relaxation in this field induced transition is unusually slow. Such slow relaxation can be explained by the formation of correlated regions inter-linked by strong dipolar interactions in these strongly polarized systems. The field induced peaks may also refer toa rise of some type of magnetic ordering which was also suggested by M(T) data. However, these results indicate the need for detailed magnetic study of geometrically frustrated systems with such combination of different magnetic ions on the single frustrated lattice which has potential to give rise to low temperature exotic states which is absent in similar systems with single magnetic element. However, other theoretical models or neutron experiments may help to further explore the origin of the observed results.

***Acknowledgements***

The authors are thankful to the Central Instrumentation Facility Centre, Indian Institute of Technology (BHU) for providing the facility of low temperature magnetic measurements. We are also thankful to Mr. R. Sahfor helping in the XAS measurements.

***Figure captions:***

**Figure 1**:Rietveld refinement of X-ray diffraction pattern collected at room temperature (300K) of all the three samples x=0.0, 0.1 and 0.2 sample. The black and red lines stand for experimental and fitted intensities respectively. The difference is shown at the bottom. The row of vertical bars representing the allowed reflections. Inset (top) : Polyhedral pictorial representation of the crystal structure, here the blue, red and green spheres are representing $Eu^{3+}$, $Ti^{4+}$ and $O^{2-}$ ions. Inset (bottom): lattice parameter and cell volume variation as a function of doping concentration.

**Figure 2:**fig (a): Fe 2p XAS spectra at $L_{2,3}$ edges recorded at 300K. Fig(b): Fe 2p XAS under +/-1T fields and the corresponding XMCD is shown in bottom panel. Fig(c) :Eu 3d XAS spectra at M4-5 edges at 300K. Fig (d) : Ti 2p XAS spectra at L2,3 edges at 300K.

**Figure 3:**Fig. (a,b and c) shows the ZFC and FC M(T) curves recorded with applied field of H=100 Oe for the samples x=0.0, 0.1 and 0.2 respectively. Fig. (d) shows the comparison of dc M(T)curves(at 100 Oe) for the samples withx=0.0, 0.1 and 0.2. Inset of Fig. (a): Combined 3D spin-wave model andCW fit. Inset of Fig. (b) and (c) show the CW fit for the range 100-300 K and combined CW-3D spin wave model fit for 2-100 K for the Fe doped samples x=0.1 and 0.2 respectively. Inset figures (a), (b) and (c) of Fig. (d) showing“high temperature series expansionof

susceptibility ($\chi$) fitting" of curve "**$\chi$T Vs 1/T**" for x=0.0, 0.1 and 0.2 samples. Sub-inset of Fig. (b) and (c) are showing a closer view of the ZFC-FC curves for x=0.1 and 0.2 respectively.

**Figure 4:**Isothermal magnetization curves as a function of magnetic fields at temperature 2K. Inset 3 showing a close view of the curve. The inset 1 and 2 are showing the M(T) curves at different applied fields.

**Figure 5: (**a) and (c) showing temperature variation of $\chi^{/}$and $\chi^{//}$ @H=0 KOe while (b) and (d) showing $\chi^{/}$and $\chi^{//}$ @H=10KOe for x=0.1 sample. Inset of (a) and (b): Enlarged view of $\chi^{/}$ (T) with H=0 KOe and 10Koe respectively, showing frequency dependence is still present in the system. Inset of (c) and (d): The temperature variation of real and imaginary parts of ac susceptibility ($\chi^{/}$and $\chi^{//}$) at different frequencies for pure $Eu_2Ti_2O_7$ at H=0 KOe.

**Figure 6:** Arrhenius law fits of frequency dependence of freezing temperatures $\boldsymbol{T_f}$ for all the three samples (x=0.0, 0.1 and 0.2). Inset: Frequency dependence of freezing temperatures for pure (x=0.0) and doped samples(x=0.1 and 0.2).

**Figure 7:** Field and frequency dependence of ac susceptibility ($\chi^{/}$and $\chi^{//}$) for x=0.1 sample. (a) and (c) showing $\chi^{/}$and $\chi^{//}$ as a function of temperature at f=300Hz at different fields. (b) and (d) showing temperature variation of $\chi^{/}$and $\chi^{//}$ at dc field H=2T with different frequencies down to 10Hz.

**Figure .8 :**(a) and (b) showing the temperature (T) variation of $\chi^{/}$and $\chi^{//}$ at dc field H=0 KOe for x=0.2 sample. Fig. (c) and (d) showing $\chi^{/}$ **(T)** and $\chi^{//}$ (T) curves with different applied dc fields at frequency 500Hz. Inset of fig (a) : Enlarged view of $\chi^{/}$ (T) @H=0 KOe

**Table1**: Structural parameters and crystallographic sites determined from Rietveld profile refinement of the powder XRD patterns for pyrochlore $\mathbf{Eu_{2-x}Fe_xTi_2O_7}$ at 300K (room temperature).

Space group: F d $\bar{3}$ m

| Parameters | X=0 | X=0.1 | X=0.2 |
|---|---|---|---|
| **Lattice constant (Å)** | **a=b=c= 10.18645(8)** | **a=b=c= 10.1722** | **a=b=c=10.1570** |
| **Cell volume (Å$^3$)** | **1056.9840** | **1052.5547** | **1047.8527** |
| **Eu site** | **16d** | **16d** | **16d** |
| **x** | **0.5** | **0.5** | **0.5** |
| **y** | **0.5** | **0.5** | **0.5** |
| **z** | **0.5** | **0.5** | **0.5** |
| **Fe site** | **16d** | **16d** | **16d** |
| **x** | **0.5** | **0.5** | **0.5** |
| **y** | **0.5** | **0.5** | **0.5** |
| **z** | **0.5** | **0.5** | **0.5** |
| **Ti site** | **16c** | **16c** | **16c** |
| **x** | **0.0** | **0.0** | **0.0** |
| **y** | **0.0** | **0.0** | **0.0** |
| **z** | **0.0** | **0.0** | **0.0** |
| **O(1) site** | **48f** | **48f** | **48f** |
| **x** | **0.31870** | **0.32495** | **0.32902** |
| **y** | **0.125** | **0.125** | **0.125** |
| **z** | **0.125** | **0.125** | **0.125** |
| **O(2) site** | **8b** | **8b** | **8b** |
| **x** | **0.375** | **0.375** | **0.375** |
| **y** | **0.375** | **0.375** | **0.375** |
| **z** | **0.375** | **0.375** | **0.375** |
| $R_{wp}$ | **29.8** | **25.4** | **25.4** |
| $R_{exp}$ | **17.01** | **16.06** | **15.75** |
| $R_{wp}/R_{exp}$ | **1.75** | **1.51** | **1.61** |
| **Chi$^2$** | **3.06** | **2.49** | **2.37** |
| $d_{Eu-Eu}$**(Å)** | **3.601454** | **3.59642(10)** | **3.59105** |
| $d_{Eu-O(1)}$**(Å)** | **2.512442(12)** | **2.50893(6)** | **2.50519** |
| $d_{Eu-O(2)}$ **(Å)** | **2.205431(10)** | **2.20235(5)** | **2.19906** |
| $d_{Ti-O(1)}$ **(Å)** | **3.79552(2)** | **3.79021(12)** | **3.78455** |
| $d_{O(1)-O(2)}$ **(Å)** | **3.02537(2)** | **3.62805(10)** | **3.62264** |
| $d_{Fe-O(1)}$ **(Å)** | **-** | **4.65409(16)** | **4.64715** |
| **<(Eu)-(O1)-(Eu)>(deg)** | **139.8122 (13)** | **139.812(6)** | **139.81218** |
| **<(Eu)-(O2)-(Eu)>(deg)** | **109.4712(9) -** | **109.471(5)** | **109.47123** |
| **<(Eu)-(O1)-(Fe)>(deg)** | **-** | **139.812(6)** | **139.81218** |
| **<(Ti)-(O1)-(Fe)>(deg)** | **-** | **106.350(5)** | **106.35051** |
| **<(Fe)-(O1)-(Fe)> (deg)** | **-** | **139.812(6)** | **139.81218** |
| **<(Ti)-(O1)-(Ti)> (deg)** | **132.9695(15)** | **132.382(6)** | **132.3825** |
| **<(O1)-(Ti)-(O1)>** | **94.5139 (8)** | **94.514 (4)** | **94.51386** |
| **<(O1)-(Ti)-(O1)>** | **85.4861(11)** | **85.486 (6)** | **85.48614** |

**Table-2**: **Showing the magnetic characteristic parameters evaluated from high temperature series expansion of susceptibility study for the temperature range 2K-5K for all three samples (x=0, 0.1, 0.2).**

| **Sample** $Eu_{2-x}Fe_xTi_2O_7$ | Curie-Weiss temperature($\Theta_{cw}$) | Effective magnetic moment ($\mu_{eff}$) | Exchange interaction energy ($J_{nn}$) | Dipolar exchange interaction energy ($D_{nn}$) |
|---|---|---|---|---|
| **X=0.0** | **-1.35K** | **0.679 $\mu_B$** | **-0.67K** | **+0.006K** |
| **X=0.1** | **-0.72K** | **1.346 $\mu_B$** | **-0.36K** | **+0.024K** |
| **X=0.2** | **-0.79K** | **1.801 $\mu_B$** | **-0.39K** | **+.045K** |

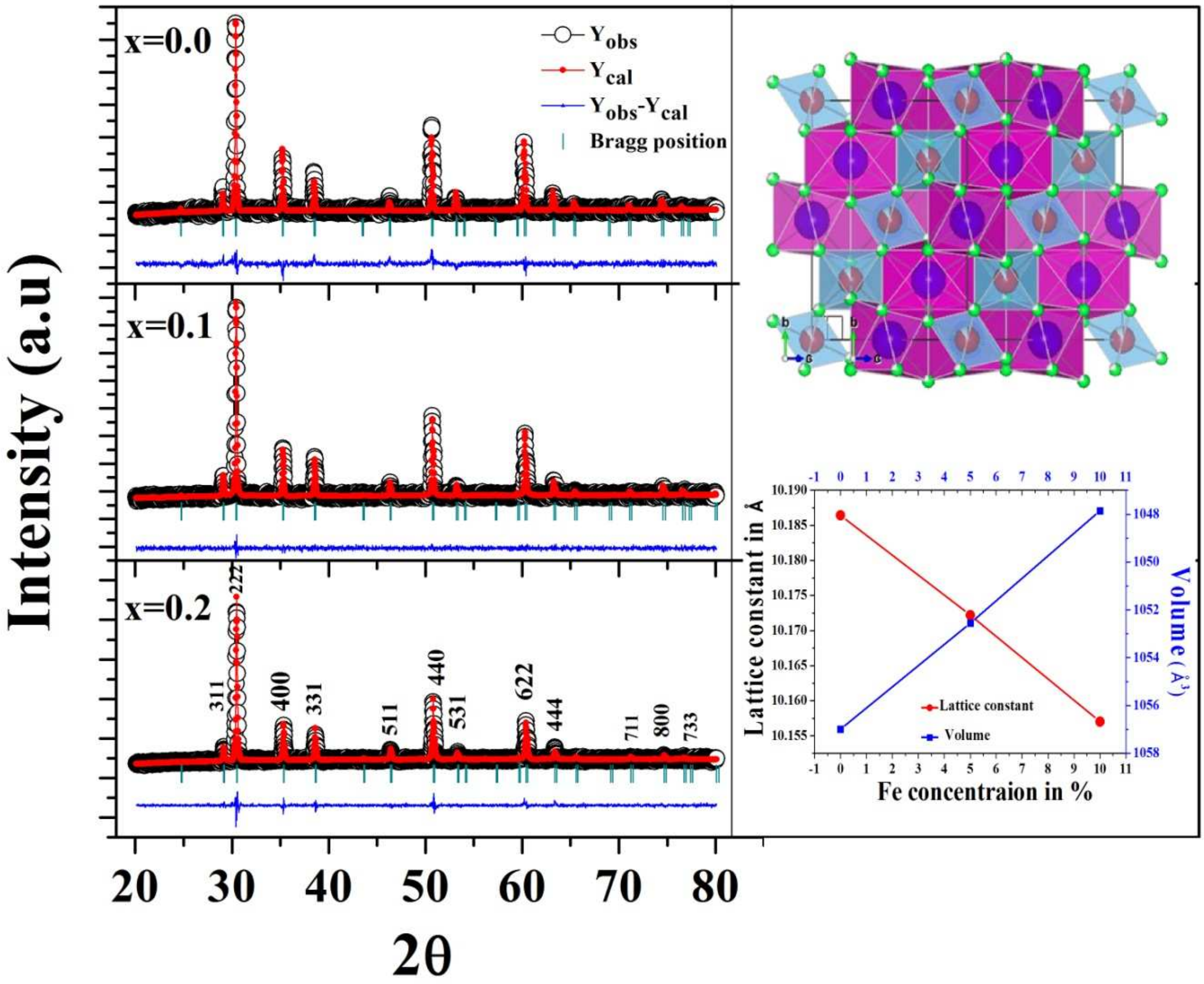

Fig.1

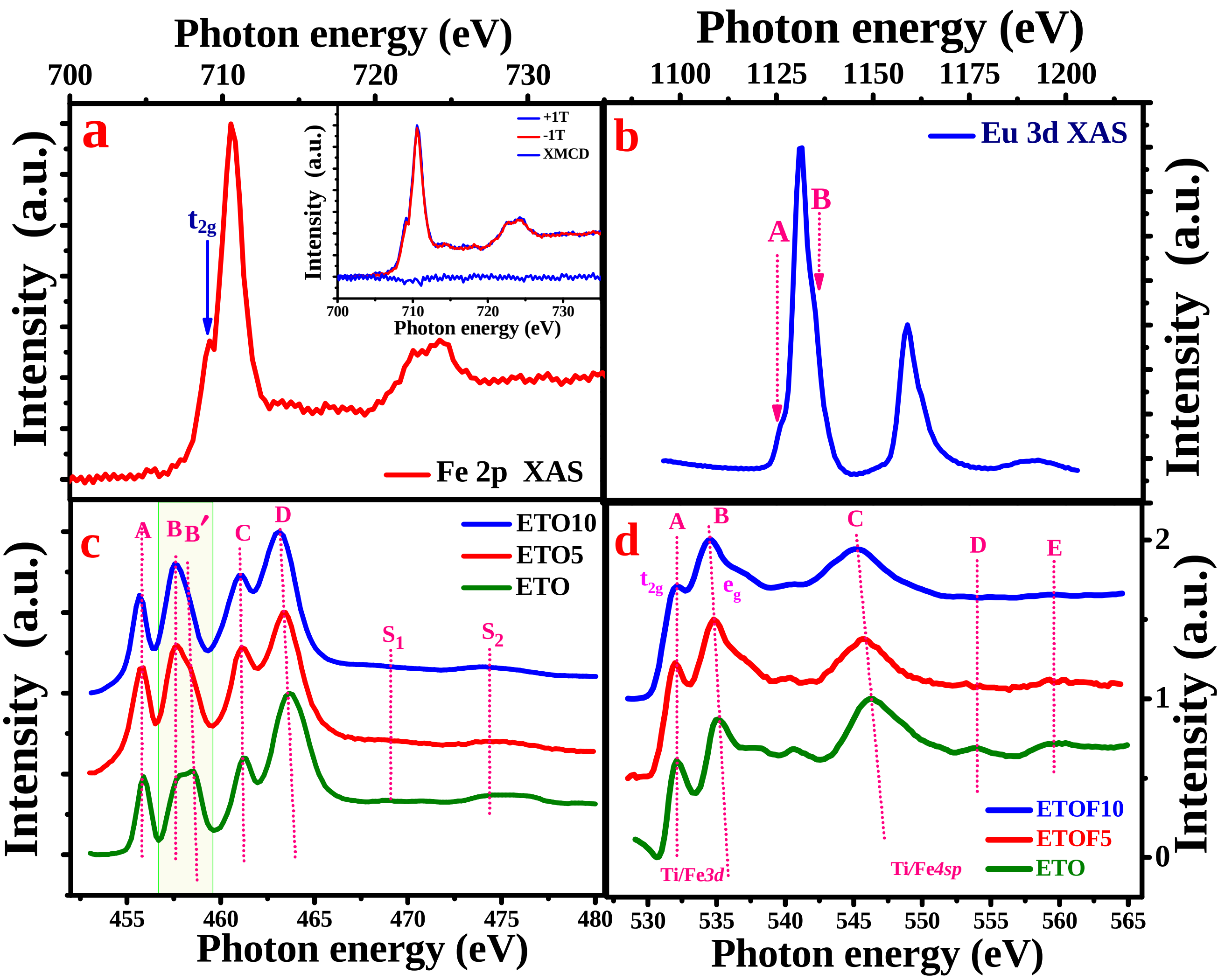

Fig.2

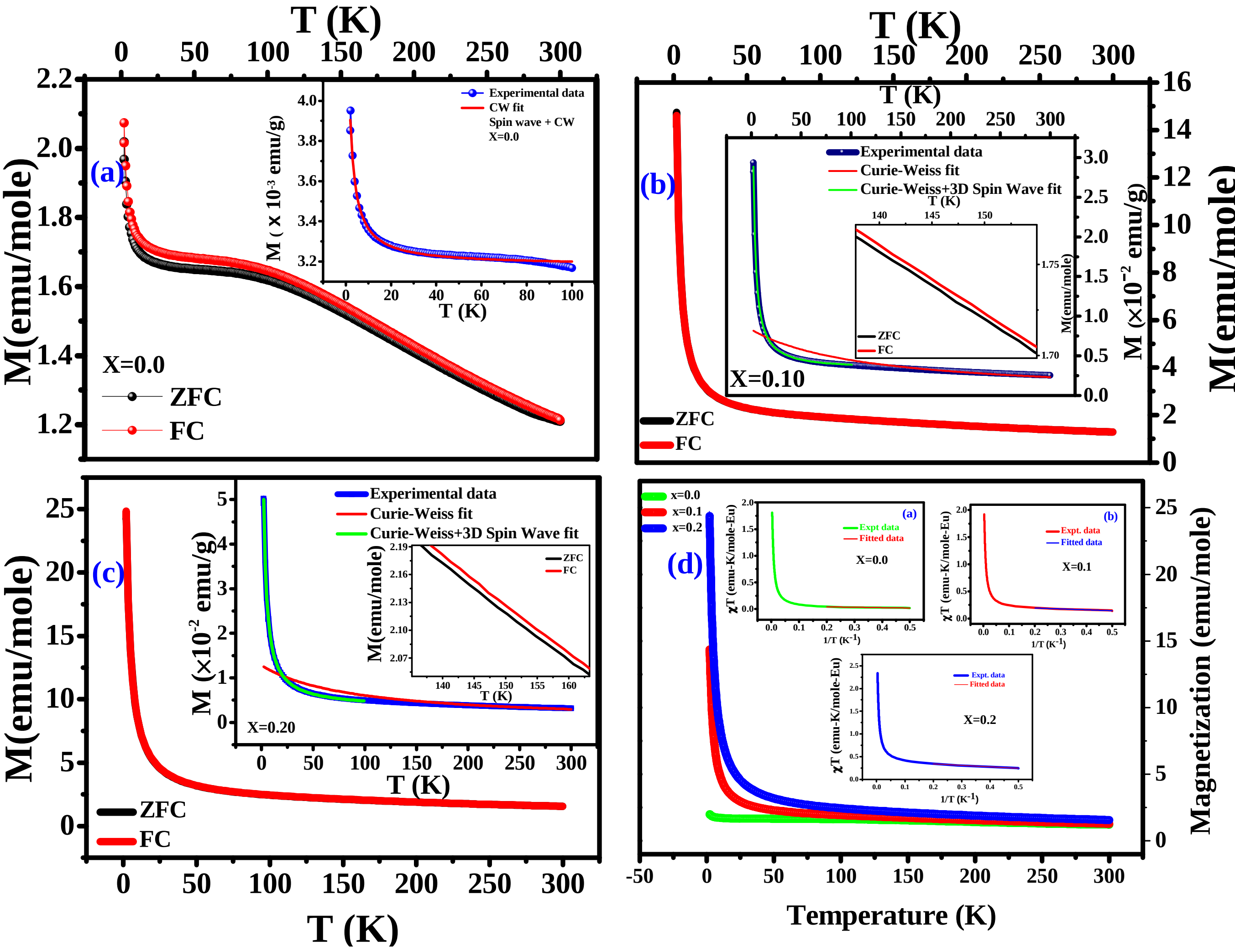

Fig3:

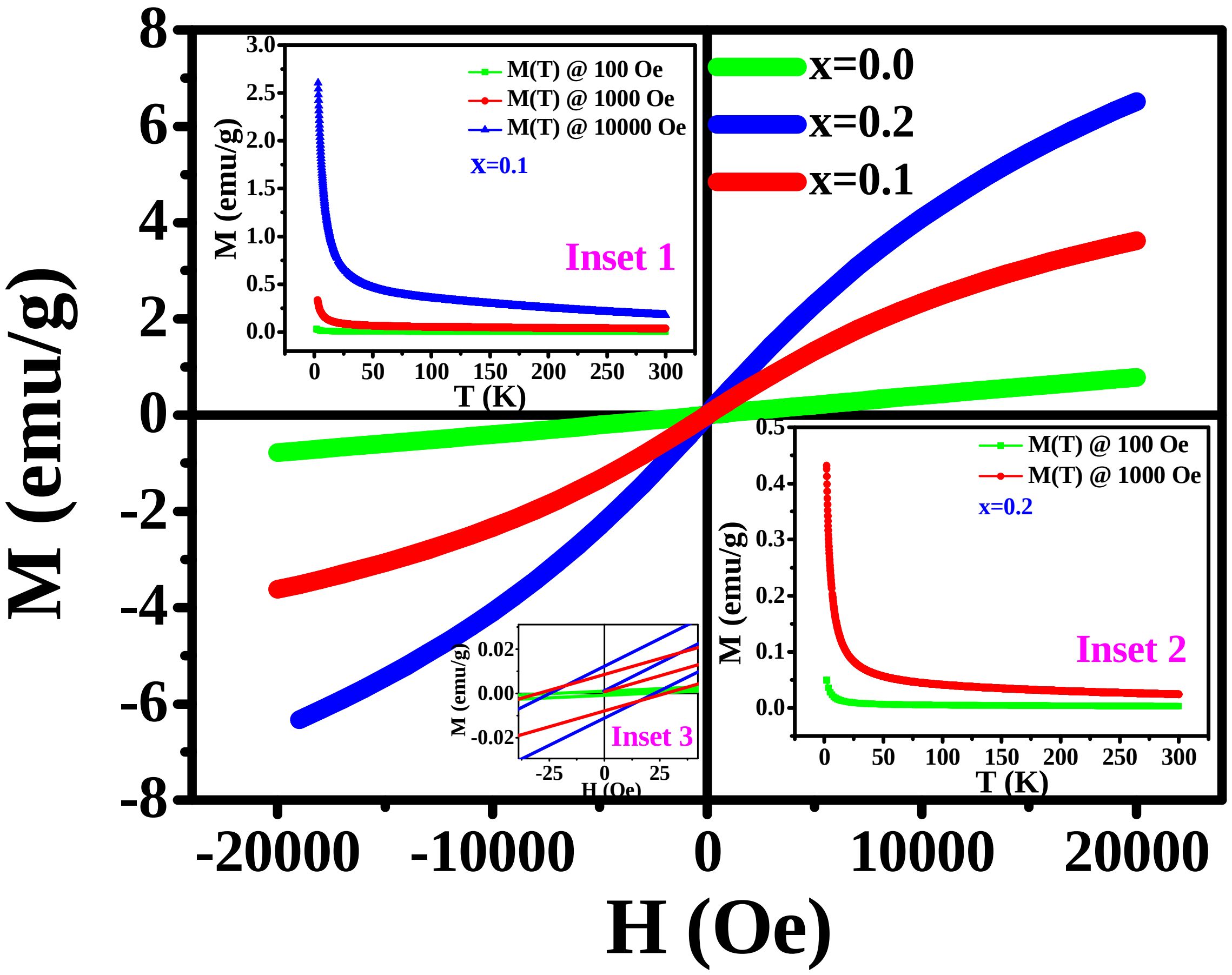

Fig. 4

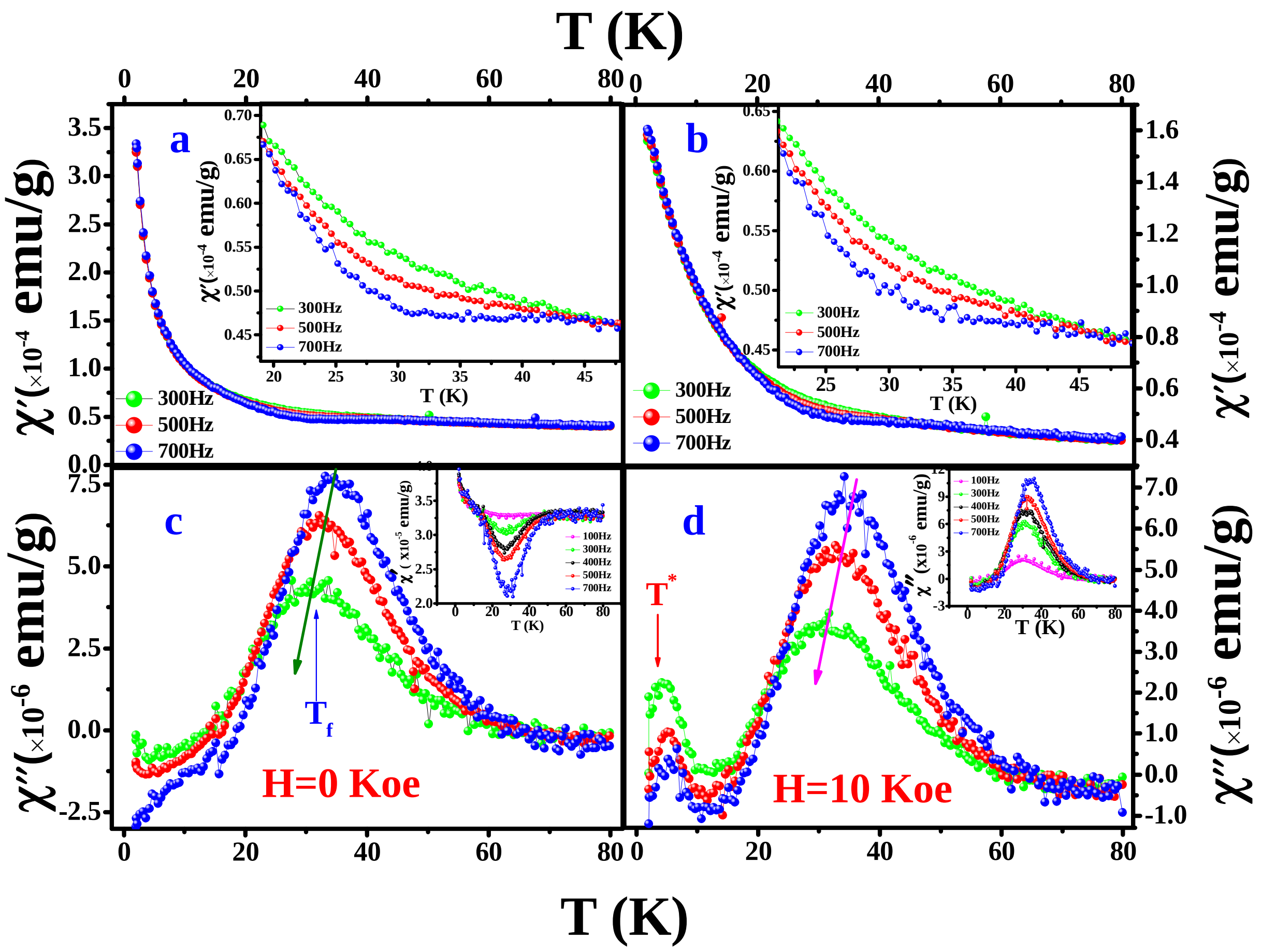

Fig.5

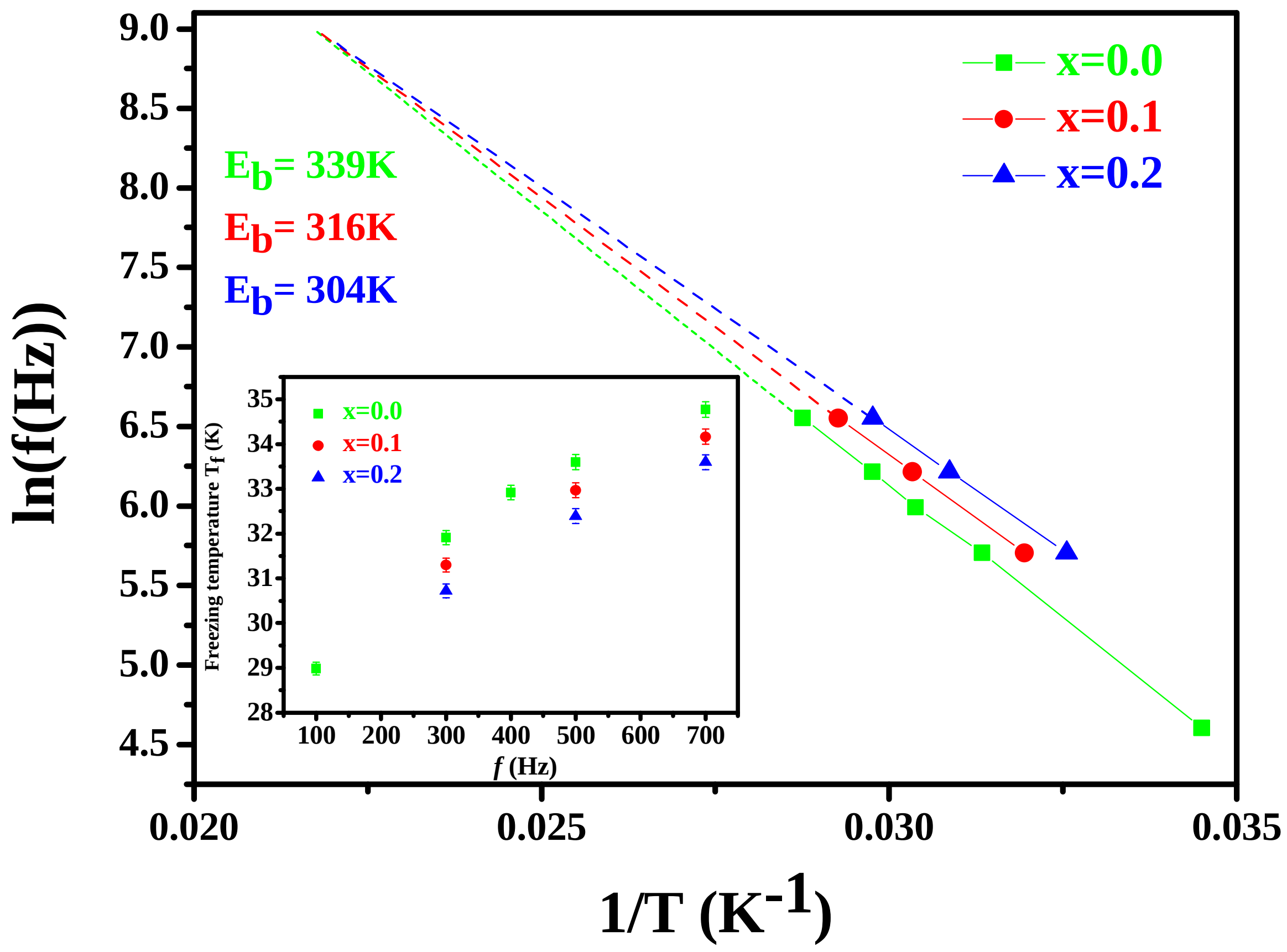

Fig. 6

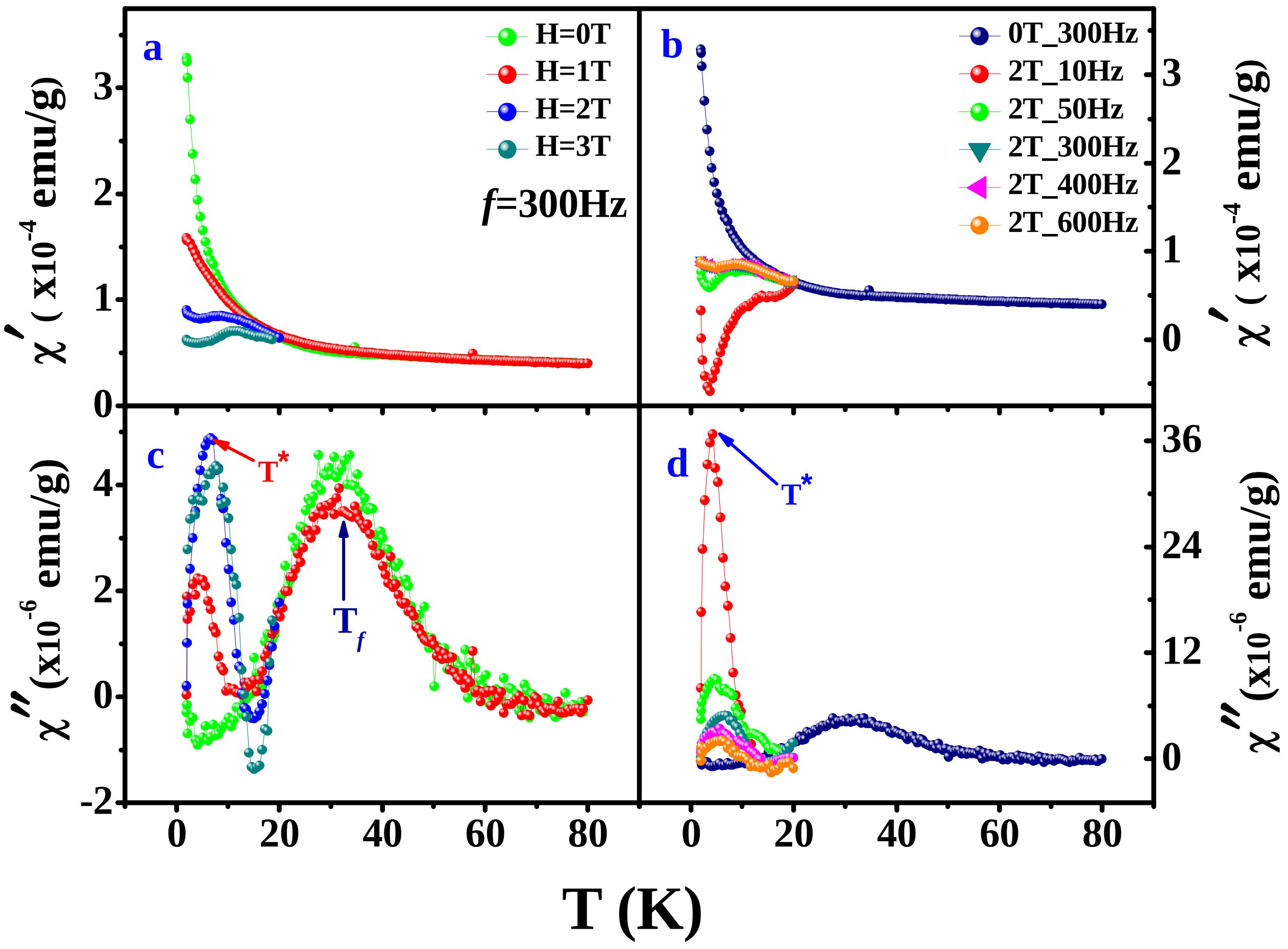

Fig .7

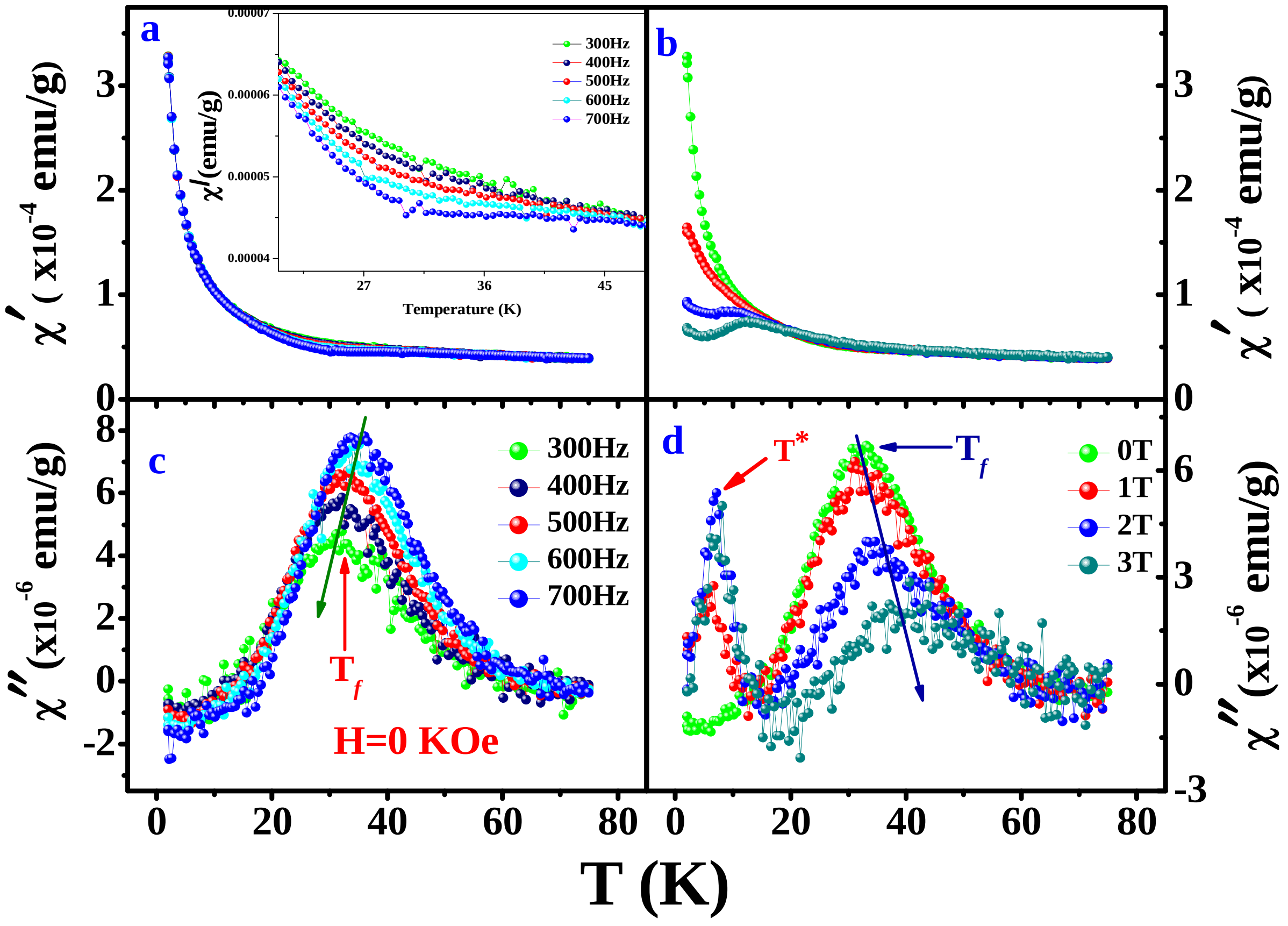

**Fig. 8**